\documentclass[10pt,a4paper]{article}
\usepackage[utf8]{inputenc}
\usepackage{amsmath}
\usepackage{amsfonts}
\usepackage{amssymb}
\usepackage{graphicx}
\usepackage{graphics}
\usepackage{cite}
\usepackage{subfig}
\usepackage{float}
\usepackage[toc,page]{appendix}
\usepackage{geometry}
\usepackage{authblk}
\usepackage{abstract}
\title{$e^+e^-$ annihilation on the stochastic background of primordial gravitational waves with BKP18}
\author[]{N. Malsawmtluangi \thanks{ Corresponding author, E-mail: \texttt{tei.naulak@uohyd.ac.in}}}
      
\affil[]{\small Department of Physics, Government Kolasib College\\ Kolasib-796081, Mizoram, India.}
\date{}

\begin{document}
\maketitle

\begin{abstract}
The evolution of inflationary primordial gravitational waves with the expansion of the universe is studied while taking into account the $e^+e^-$ annihilation which leads to step correction in the spectrum of primordial gravitational waves which is $\sim 10\%$ in the amplitude and $\sim 25\%$ in the energy density. The amplitudes and energy densities are studied for the $\alpha$-attractor inflation models which are in agreement with the recent BICEP/Keck and Planck 2018 joint result. The energy densities are well within the big bang nucleosynthesis bound, but the spectra for these models are very much below the detection range of the currently operating and upcoming detectors.
\end{abstract}

\section{Introduction}
\label{intro}
Primordial gravitational waves generated during inflation have traversed the universe throughout its expansion and evolution. As such, primordial gravitational waves carry information about the very early universe and its subsequent evolution that are otherwise difficult or even impossible to be probed through other means. They are believed to form a stochastic background with standing wave pattern over all frequency with spectrum and energy depending on the evolutionary stages of the universe. Hence they are present everywhere at anytime at all frequencies \cite{as1,lp1,yz1,mg}. However, as their wavelengths have also stretched along with the expansion of universe, their energies have diminished significantly and are the most difficult form of gravitational waves to detect, whether directly or indirectly. 

If detected, the stochastic gravitational waves would provide an evidence of and information on the physical processes taking place during the very early stages of the universe. For primordial gravitational waves generated during inflation, the amplitude is directly related to the energy scale during inflation. The lowest frequency mode corresponds to the present-day horizon size while the highest frequency mode of the spectral energy density corresponds to the energy scale of reheating, a process which follows the decay of the inflaton, the scalar field that drives the inflation, after the end of the inflationary process. Thus the spectrum of primordial gravitational waves can provide information on the very early processes of the universe.

The spectrum of the primordial gravitational waves is also believed to be affected by the changes in the physical conditions in the universe. For  instance, the $e^+e^-$ annihilation that occurs at frequency $\sim 10^{-10}$ Hz causes a change in the number of relativistic degrees of freedom \cite{as2,djs,djs2,ywk,mmg}. This further affects the expansion behavior of the universe and subsequently leads to a step in the spectrum of primordial gravitational waves. This is because the growth rate of the Hubble radius gets reduced during the process of $e^+e^-$ annihilation which leads to the change in the rate of horizon re-entry of the modes at that instant, which is within the radiation dominated era and a step in the gravitational wave spectrum appears at the frequency of the order of Hubble rate at that instant. This step would be about $\sim 10 \%$ in the amplitude and about $\sim 20 \%$ in the energy spectrum \cite{djs,ywk,swy}.

The primordial gravitational waves produced during the inflationary period are believed to exist in a specific quantum state, known as squeezed vacuum state \cite{lp2,as3,sk,amn,lp3,lp4,lp5,np}. This is due to the fact that the inflationary quantum vacuum field fluctuations generated non-zero variance for quantum fluctuations which, due to parametric amplification, transformed the initial vacuum state into a quantum state with multiple particles, a state known as the squeezed vacuum state. Due to this process, the phase variance of the wave mode is being squeezed while there is increase in the variance of its amplitude at the same time so that the uncertainty product is being held. 

While the direct detectability of stochastic background of primordial gravitational waves has been questioned \cite{que}, recently there are studies on their potential detectability via quantum noise. Quantum mechanical treatment of gravitational field is shown to be able to induce fluctuations or noise in the lengths of the arms of gravitational wave detectors where the characteristics of the noise depend on the quantum state of the gravitational field and while such noise is very small for coherent states, it can be greatly enhanced especially in squeezed states by an exponential of the squeezing parameter and hence, potentially detectable \cite{qn1,qn2,qn3,qn4,qn5}. Therefore, it is interesting to study the stochastic signals of the inflationary gravitational waves in wake of these recent developments.

In this paper, we calculated the relevant parameters for the stochastic background of primordial gravitational waves including the $e^+e^-$ annihilation stage for the $\alpha$-attractor inflation models which are in good agreement with the 2021 release of the BICEP/Keck and 2018 Planck joint data (BKP18) \cite{bkp}. The paper is organized as follows. In section \ref{sec2}, we reviewed the epochs of expansion of the universe starting from inflation upto the present accelerating stage including the $e^+e^-$ annihilation. In section \ref{sec3}, we obtained the expression for the amplitude of the primordial gravitational waves with the associated parameters \cite{lp1,lp5,np}. In section \ref{sec4}, we provided relevant parameters to obtain the spectra of the gravitational waves normalized with the BKP18 data. We also briefly introduced the $\alpha$-attractor models and obtained the inflationary and reheating indices for these models. We also showed the resultant amplitude and energy spectrum and compared them with the big bang nucleosynthesis bound \cite{bbn1,bbn2,bbn3,bbn4}. We also checked the detectability of the resultant amplitude with several ongoing and proposed gravitational wave detectors like Advanced LIGO, Advanced Virgo, LIGO A+, KAGRA, Einstein Telescope and Cosmic Explorer \cite{sc,alg,adv1,adv2,apl,kg1,kg2,et,ce1,ce2,g23,ks}. We presented our conclusions in section \ref{sec5}. In Appendix A, we gave a brief account of the physics of particles before and after the $e^+e^-$ annihilation, concentrating on the effective relativistic number of species \cite{ywk,mmg}. In Appendix B, we presented the growth of the wave mode $h_k(\varsigma)$. In Appendix C, we showed how to obtain the value of reheating parameter when $e^+e^-$ annihilation stage is not considered explicitly in the evolutionary stages of the universe for comparison.

\section{Epochs of expansion of the universe}
\label{sec2}
The universe has undergone several stages of evolution throughout its expansion. The successive epochs of the evolution can be characterized by the scale factor $a$ in power-law form \cite{yz1,swy}.

The power-law evolution of the scale factor with proper time $t$ can be obtained directly from the conservation equation and the Friedmann equations for the flat universe as $a \propto t^{2/3(1+w)}$, then in terms of conformal time $d\varsigma=dt/a$, one can obtain,
\begin{equation}\label{a1}
a(\varsigma) = l_0 |\varsigma|^{1+\beta},~~~~~~~~~~-\infty \leqslant \varsigma \leqslant \varsigma_1,
\end{equation}
where $l_0$ and $\beta$ are arbitrary constants, $\varsigma_1 < 0$ and $\beta + 1 < 0$. The inflationary index $\beta$ is related to the effective equation of state $w= p/ \epsilon$ of matter governing the inflationary stage, $p$ being the effective pressure of matter and $\epsilon$ being the energy density as, $w = (1-\beta)/3(1+\beta)$ and $w$ varies from $-1/3$ to $-\infty$ for $-\infty < \beta < -1$ \cite{lp7,jm}. The case $\beta = -2$ corresponds to the de Sitter state.

The reheating stage:
\begin{equation}\label{a2}
a(\varsigma) = a_z |\varsigma - \varsigma_p|^{1+\beta_s},~~~~~~~~~~\varsigma_1 \leqslant \varsigma \leqslant \varsigma_s,
\end{equation}
where $\beta_s$ is the reheating parameter. The subscript $z$ is used to denote the reheating stage.

The radiation-dominated stage:
\begin{equation}\label{a3}
a(\varsigma) = a_e (\varsigma - \varsigma_e),~~~~~~~~~~\varsigma_s \leqslant \varsigma \leqslant \varsigma_y,
\end{equation}
where the subscript $e$ denotes this stage.

The $e^+e^-$ annihilation:
\begin{equation}\label{a4}
a(\varsigma) = a_v (\varsigma - \varsigma_v)^{1+v},~~~~~~~~~~\varsigma_y \leqslant \varsigma \leqslant \varsigma_z,
\end{equation}
where the subscript $v$ denotes the period for $e^+e^-$ annihilation which starts from $\varsigma_y$ and ends at $\varsigma_z$. This phenomenon occurs during radiation-dominated period. 

Throughout this paper, the power index $v=0.0616$ will be used for calculation purposes \cite{swy}. See also Appendix \ref{app} for details.

After $e^+e^-$ annihilation and before matter domination:
\begin{equation}\label{a5}
a(\varsigma) = a_g(\varsigma-\varsigma_g),~~~~~~~~~~\varsigma_z \leqslant \varsigma \leqslant \varsigma_2,
\end{equation}
where the subscript $g$ is used to denote this period which is still in the radiation dominated era.

The matter-dominated stage:
\begin{equation}\label{a6}
a(\varsigma) = a_m (\varsigma - \varsigma_m)^2,~~~~~~~~~~ \varsigma_2 \leqslant \varsigma \leqslant \varsigma_E,
\end{equation}
where the subscript $m$ denotes this stage.

The accelerating stage upto the present time:
\begin{equation}\label{a7}
a(\varsigma) = l_H |\varsigma - \varsigma_a|^{-1},~~~~~~~~~~ \varsigma_E \leqslant \varsigma \leqslant \varsigma_H,
\end{equation}
where $l_H$ is the present day Hubble radius and  is taken as,
\begin{equation}
l_H = \left(\frac{a^2}{a'}\right) = \frac{1}{H}.
\end{equation}
 The power index in eq.\eqref{a7} is often given by $\gamma$ which depends on the dark energy \cite{yz2}. Even though the matter component also exists, since the dark matter is dominant, we shall take the approximation $\gamma=1$ to the current acceleration behavior throughout this paper for calculation purposes.

The conformal time instances $\varsigma_1$, $\varsigma_s$, $\varsigma_y$, $\varsigma_z$, $\varsigma_2$, $\varsigma_E$, $\varsigma_H$ denote the various successive evolutionary stages of the universe. By choosing the overall normalization $|\varsigma_H-\varsigma_a|=1$, the continuous joining of the functions $a(\varsigma)$ and $a'(\varsigma)$, where $'$ is derivative with respect to conformal time, at these points of transition provide the link between the conformal time instances in eqs.\eqref{a1}-\eqref{a7} as:
\begin{eqnarray}
\varsigma_a - \varsigma_E &=& \xi_E, \nonumber \\
\varsigma_E - \varsigma_m &=& 2\xi_E, \nonumber \\
\varsigma_2 - \varsigma_m &=& 2\xi_E\xi_2^{-1/2}, \nonumber \\
\varsigma_2 - \varsigma_g &=& \xi_E \xi_2^{-1/2}, \nonumber \\
\varsigma_z - \varsigma_g &=& \xi_E \xi_2^{-1/2} \xi_z^{-1}, \nonumber \\
\varsigma_z - \varsigma_v &=& (1+v) \xi_E \xi_2^{-1/2} \xi_z^{-1},  \label{vrs} \\
\varsigma_y - \varsigma_v &=& (1+v) \xi_E \xi_2^{-1/2} \xi_z^{-1} \xi_y^{-(1/1+v)}, \nonumber \\
\varsigma_y - \varsigma_e &=& \xi_E \xi_2^{-1/2} \xi_z^{-1} \xi_y^{-(1/1+v)}, \nonumber \\
\varsigma_s - \varsigma_e &=& \xi_E \xi_2^{-1/2} \xi_z^{-1} \xi_y^{-(1/1+v)} \xi_s^{-1}, \nonumber \\
\varsigma_s - \varsigma_p &=& |1+\beta_s| \xi_E \xi_2^{-1/2} \xi_z^{-1} \xi_y^{-(1/1+v)} \xi_s^{-1}, \nonumber \\
\varsigma_1 - \varsigma_p &=& |1+\beta_s| \xi_E \xi_2^{-1/2} \xi_z^{-1} \xi_y^{-(1/1+v)} \xi_s^{-1} \xi_1^{-(1/1+\beta_s)}, \nonumber \\
\varsigma_1 &=& (1+\beta) \xi_E \xi_2^{-1/2} \xi_z^{-1} \xi_y^{-(1/1+v)} \xi_s^{-1} \xi_1^{-(1/1+\beta_s)}, \nonumber
\end{eqnarray}
and the arbitrary constants as:
\begin{eqnarray}\label{ars}
a_m &=& \frac{l_H}{4}\xi_E^{-3}, \nonumber \\
a_g &=& l_H \xi_E^{-2} \xi_2^{-1/2}, \nonumber \\
a_v &=& l_H (1+v)^{-(1+v)} \xi_E^{-(2+v)} \xi_2^{(v-1)/2}\xi_z^v, \nonumber \\
a_e &=& l_H \xi_E^{-2} \xi_2^{-1/2} \xi_y^{-(v/1+v)}, \nonumber \\
a_z &=& l_H |1+\beta_s|^{-(1+\beta_s)} \xi_E^{-(2+\beta_s)} \xi_2^{(\beta_s-1)/2}  \xi_y^{(\beta_s-v)/(1+v)} \xi_s^{\beta_s} \xi_z^{\beta_s}, \nonumber \\
l_0 &=& l_Hb \xi_E^{-(2+\beta)} \xi_2^{(\beta-1)/2} \xi_z^\beta \xi_y^{(\beta-v)/(1+v)}  \xi_s^\beta \xi_1^{(\beta - \beta_s)/(1+\beta_s)},
\end{eqnarray}
where, $b= | 1+\beta|^{-(1+\beta)}$ and,
\begin{eqnarray}
&& \xi_E = \frac{a(\varsigma_H)}{a(\varsigma_E)},~~~~~\xi_2 = \frac{a(\varsigma_E)}{a(\varsigma_2)},~~~~~\xi_z = \frac{a(\varsigma_2)}{a(\varsigma_z)}, \nonumber \\
&& \xi_y = \frac{a(\varsigma_z)}{a(\varsigma_y)},~~~~~~\xi_s = \frac{a(\varsigma_y)}{a(\varsigma_s)},~~~~~\xi_1 = \frac{a(\varsigma_s)}{a(\varsigma_1)}.
\end{eqnarray}

\section{Primordial gravitational waves}
\label{sec3}
The perturbed metric of a flat FLRW universe in the presence of gravitational waves is,
\begin{equation}
dS^2 = a^2(\varsigma)[-d\varsigma^2 + (\delta_{ij}+h_{ij})dx^idx^j],
\end{equation}
where $|h_{ij}|\ll\delta_{ij}$ is a transverse-traceless perturbation of space-time, $\partial_ih^{ij} = 0$, $\delta^{ij}h_{ij}=0$, $\delta_{ij}$ being the flat space metric.

The gravitational wave field $h_{ij}(\textbf{x},\varsigma)$ can be expanded over spatial Fourier harmonics $e^{\pm i\textbf{k}.\textbf{x}}$ as
\begin{eqnarray}
h_{ij}(\textbf{x},\varsigma) &=& \frac{D}{(2\pi)^\frac{3}{2}}\int_{-\infty}^{+\infty}\frac{d^3\textbf{k}}{\sqrt{2k}}\sum_{p=1}^2 [h_k^{(p)}(\varsigma) c_k^{(p)} \varepsilon_{ij}^{(p)}(\textbf{k})e^{i\textbf{k}.\textbf{x}}  + h_k^{(p) \ast}(\varsigma)c_k^{(p) \dagger} \varepsilon_{ij}^{(p) \ast}(\textbf{k})e^{-i\textbf{k}.\textbf{x}}],
\end{eqnarray}
where $D=\sqrt{16\pi}l_{pl}$ is the normalization constant, $l_{pl}=\sqrt{G}$ is Planck's length. The wave number $k$ is related to the wave vector $\textbf{k}$ as, $k=(\delta_{ij}k^ik^j)^\frac{1}{2}$ and is related to wavelength $\lambda$ by $\lambda = 2\pi a/k$.

The two linear polarization states $\varepsilon_{ij}^{(p)}$, where $p=1,2$, are symmetric and transverse-traceless and satisfy the conditions
$\varepsilon_{ij}^{(p)}\delta^{ij}=0$, $\varepsilon_{ij}^{(p)}k^i=0$, $\varepsilon_{ij}^{(p)}\varepsilon^{(p') ij} = 2\delta_{pp'}$, $\varepsilon_{ij}^{(p)}(\textbf{-k})=\varepsilon_{ij}^{(p)}(\textbf{k})$.
Since the contributions from both these polarizations are same, the superscript $p$ is dropped from here onward for convenience.

The creation and annihilation operators $c_k^{\dagger}$ and $c_k$ satisfy the relationships $[c_k,c_{k'}^{ \dagger}]=\delta^3(k-k')$, $[c_k,c_{k'}]=[c_k^{\dagger},c_{k'}^{ \dagger}]=0$ and $c_k |0\rangle = 0$, where $|0\rangle$ is the initial vacuum state.

The wave equation of the primordial gravitational waves in the flat FLRW universe in terms of the mode $h_k$ can be written as,
\begin{equation}
h''_k (\varsigma) + 2\frac{a'}{a} h'_k (\varsigma) + k^2h_k (\varsigma) = 0.
\end{equation}
The function $h_k(\varsigma)$ is a complex time-dependent function which represents the time-evolution of all $\textbf{k}$ belonging to a given $k$ and can be rescaled in terms of mode function for $k$ as
\begin{equation}\label{mf}
h_k(\varsigma) a(\varsigma)= \mu_k(\varsigma),
\end{equation}
where the mode functions can have the following form
\begin{equation}\label{mff}
\mu_k(\varsigma) = u_k(\varsigma) + v_k^{\ast}(\varsigma),
\end{equation}
where $u(k)$ and $v(k)$ are complex functions and can be represented in terms of three real functions - the squeezing parameter $r_k$, squeezing angle $\phi_k$ and the rotation angle $\theta_k$ as \cite{lp5},
\begin{eqnarray} \label{y}
u_k &=& e^{i \theta_k} \cosh r_k, \nonumber \\
v_k &=& e^{-i(\theta_k -2\phi_k)} \sinh r_k.
\end{eqnarray}
In the course of evolution of the universe, the complex functions $u_k(\varsigma) + v_k^\ast (\varsigma)$ from eq.\eqref{mff} become practically real such that $h_k(\varsigma)a(\varsigma) = h_k^\ast (\varsigma) a(\varsigma) = e^{r_k}\cos \phi_k$ \cite{lp1} (see also Appendix \ref{appb}). These two complex functions $u_k$ and $v_k$ satisfy the equations,
\begin{eqnarray}
i\frac{du_k}{d\varsigma} &=& ku_k + i\frac{a'}{a} v^{\ast}_k, \nonumber \\
 i\frac{dv_k}{d\varsigma} &=& kv_k + i\frac{a'}{a} u^{\ast}_k,
\end{eqnarray}
which lead to the equations governing the three real functions as:
\begin{eqnarray}\label{aaa}
r'_k &=&\frac{a'}{a}\cos 2\phi_k, \nonumber \\  
\phi'_k &=& -k-\frac{a'}{a}\sin 2\phi_k \coth 2r_k,\\ 
\theta'_k &=& -k-\frac{a'}{a} \sin 2\phi_k \tanh r_k. \nonumber
\end{eqnarray}
The power spectrum of the gravitational waves is defined by two-point correlation function of the field $h_{ij}$,
\begin{eqnarray}\label{2pt}
\langle0|h_{ij}(\textbf{x},\varsigma)h^{ij}(\textbf{x},\varsigma)|0\rangle &=& \frac{D^2}{2\pi^2}\int_{0}^{\infty}k^2 |h_k(\varsigma)|^2\frac{dk}{k}  = \int_{0}^{\infty} h^2(k,\varsigma) \frac{dk}{k},
\end{eqnarray}
where $h^2(k,\varsigma)=\frac{D^2}{2\pi^2}k^2|h_k(\varsigma)|^2$ gives the mean-square value of the gravitational waves with interval $k$,
\begin{equation}\label{2eqs}
h^2(k,\varsigma) =\frac{1}{2}|h(k,\varsigma)|^2, ~~~~{\rm and,}~~~~|h(k,\varsigma)|=\frac{D}{\pi}k|h_k(\varsigma)|.
\end{equation}
Using eqs.\eqref{mff} and \eqref{y} in eq.\eqref{mf}, we get
\begin{equation}\label{axx}
h_k^2 (\varsigma) = \frac{1}{a^2(\varsigma)}(\cosh 2r_k + \cos 2 \phi_k \sinh 2 r_k).
\end{equation}
Then, using eq.\eqref{2eqs}, we get
\begin{equation}
h(k,\varsigma) = \frac{4l_{pl}}{\sqrt{\pi}}\frac{k}{a(\varsigma)}(1+2\sinh^2r_k + \sinh2r_k\cos 2\phi_k)^{1/2}.
\end{equation}
The wave number corresponding to the current Hubble radius is,
\begin{equation}
k_H = \frac{2\pi a(\varsigma_H)}{l_H}.
\end{equation}
Then the amplitude of the primordial gravitational waves for the entire frequency range, from the end of inflationary stage to the current accelerating stage can be expressed by,
\begin{eqnarray}\label{hkbe}
h(k,\varsigma_H) &=& 8\sqrt{\pi}\left(\frac{l_{pl}}{l_H}\right) \left(\frac{k}{k_H}\right) (1 + 2\sinh^2r_k + \sinh 2r_k \cos 2 \phi_k)^{1/2}.
\end{eqnarray}

\subsection{Determining the parameters}
\label{3.1}
The squeezing parameter and squeezing angle are growing functions of time. They evolve along with the evolution of each stage of expansion. 
In the adiabatic regime for a given mode, the wavelength is shorter than the Hubble radius, therefore $k$ is dominant. Thus the squeezing angle can be given by,
\begin{equation}\label{ph1}
\phi_k = -k(\varsigma + \varsigma_k) = -\frac{k}{a(\varsigma)}\left(1+\frac{a_{\ast}(\varsigma)}{a(\varsigma_k)}\right),
\end{equation}
where $\varsigma_k$ is constant and the evaluation is done at present time where $a(\varsigma) \propto \varsigma^{-1}$. In order to find $a_{\ast}(\varsigma)/a(\varsigma_k)$, we consider $a(\varsigma_k)$ at $a(\varsigma_H)$ and $a_{\ast}(\varsigma)$ as the conformal time at the beginning of time range.

In the long wavelength regime, $k$ can be neglected, and the squeezing angle becomes,
\begin{equation}\label{ph2}
\phi_k \propto \tan^{-1} \left(\frac{1}{a^2(\varsigma)}\right),~~~~~k \leqslant k_H.
\end{equation}
Using eqs.\eqref{ph1} and \eqref{ph2}, the squeezing angle for each frequency intervals are calculated as \cite{np},
\begin{eqnarray}
\phi_k &=& -k \left[1+ \left(\frac{k_s}{k}\right)^{\beta_s} \left(\frac{k_z}{k_y}\right)^v \left(\frac{k_H}{k}\right)\left(\frac{k_H}{k_2}\right)\left(\frac{k_E}{k_H}\right)^3\right],  ~~~ k_s \leqslant k \leqslant k_1, \\
\phi_k &=& -k \left[1+ \left(\frac{k_z}{k_y}\right)^v \left(\frac{k_H}{k}\right)\left(\frac{k_H}{k_2}\right)\left(\frac{k_E}{k_H}\right)^3\right],   ~~~ k_y \leqslant k \leqslant k_s, \\
\phi_k &=& -k \left[1+ \left(\frac{k_z}{k}\right)^v\left(\frac{k_H}{k}\right) \left(\frac{k_H}{k_2}\right)\left(\frac{k_E}{k_H}\right)^3\right],   ~~~ k_z \leqslant k \leqslant k_y, \\
\phi_k &=& -k \left[1+ \left(\frac{k_H}{k}\right) \left(\frac{k_H}{k_2}\right)\left(\frac{k_E}{k_H}\right)^3\right],  ~~~ k_2 \leqslant k \leqslant k_z, \\
\phi_k &=& -k\left[1+\left(\frac{k_H}{k}\right)^2\left(\frac{k_E}{k_H}\right)^3\right], ~~k_H \leqslant k \leqslant k_2,\\
\phi_k &=& \tan^{-1}(k_H^2),~~k_E \leqslant k \leqslant k_H,\\
\phi_k &=& \tan^{-1}(k_E^2), ~~ k \leqslant k_E.
\end{eqnarray}
For each of these ranges, the squeezing angle $\cos \phi_k =1$ such that $r_k'=a'/a$  in eq.\eqref{aaa}. Hence, after integrating this, the frequency dependent squeezing parameter $r_k$ grows as,
\begin{equation}
r_k \approx \ln \frac{a_{\ast\ast}(k)}{a_{\ast}(k)},
\end{equation}
where $a_{\ast}$ is the value of $a(\varsigma)$ at $\varsigma_{\ast}$, the start of conformal time of each stage, i.e., the higher frequency end of the range since $\varsigma \propto 1/k$ and $a_{\ast\ast}$ denotes $a(\varsigma)$ at $\varsigma_{\ast\ast}$, the end of conformal time of the stage, i.e., the lower frequency end of range.

For the high frequency mode $k \geqslant k_1$, $a_{\ast} = a_{\ast\ast} = a(\varsigma_1)$ which gives $r_k = 0$. Thus the high frequency modes $k > k_1$ are not in the amplifying regime. The amplifying regime starts from $k=k_1$.

In order to find the present-day value of $r_k$, one needs to find the ratio $a_{\ast\ast}(k)/a_{\ast}(k)$. For every given wave number $k$, the quantity $a_{\ast \ast}$ is determined by the condition $\lambda (\varsigma_{\ast \ast}) = l (\varsigma_{\ast \ast})$, whereas  the quantity $a_\ast$ is determined by the condition $\lambda (\varsigma_\ast) = l (\varsigma_\ast)$, where $l (\varsigma_{\ast \ast})$ and $l (\varsigma_\ast)$ are the Hubble radius $l(\varsigma) = a^2/a'$ at $\varsigma_{\ast \ast}$ and $\varsigma_\ast$ respectively and $\lambda (\varsigma_{\ast \ast})$ and $\lambda (\varsigma_\ast)$ are the corresponding wavelengths, $\lambda = 2\pi a(\varsigma)/k$. In this manner, we find
\begin{eqnarray}\label{vs2}
&& \xi_1 = \left(\frac{k_1}{k_s}\right)^{1+\beta_s},~~~~~~~\xi_s = \frac{k_s}{k_y}, \nonumber \\
&& \xi_y = \left(\frac{k_y}{k_z}\right)^{1+v}, ~~~~~~~~~ \xi_z = \frac{k_z}{k_2}, \\
&& \xi_2 = \left(\frac{k_2}{k_E}\right)^2, ~~~~~~~~~~\xi_E = \left(\frac{k_E}{k_H}\right)^{-1}. \nonumber
\end{eqnarray}
Then, the squeezing parameter $r_k$ can be calculated in descending order of wave number by finding $a_{\ast \ast}(k)/a_\ast(k)$ for each range and then multiplying it by the previous range's $e^{r_k}$ \cite{lp1}. The squeezing parameter for each frequency range for short-wavelength regime, $k\geqslant k_H$ are then found to be,
\begin{eqnarray}
r_k &=& \ln \left(\frac{k}{k_1}\right)^{\beta - \beta_s}, ~~~ k_s \leqslant k \leqslant k_1, \label{rk1} \\
r_k &=& \ln \Big{[}\left(\frac{k}{k_s}\right)^{\beta}\left(\frac{k_s}{k_1}\right)^{\beta - \beta_s}\Big{]}, ~~k_y \leqslant k \leqslant k_s, \label{rk2}\\
r_k &=& \ln \Big{[}\left(\frac{k_y}{k}\right)^v\left(\frac{k}{k_s}\right)^\beta \left(\frac{k_s}{k_1}\right)^{\beta-\beta_s}\Big{]},   ~~~ k_z \leqslant k \leqslant k_y, \label{rk3} \\
r_k &=& \ln \Big{[}\left(\frac{k_y}{k_z}\right)^v\left(\frac{k}{k_s}\right)^\beta \left(\frac{k_s}{k_1}\right)^{\beta-\beta_s}\Big{]},    ~~~ k_2 \leqslant k \leqslant k_z, \label{rk4} \\
r_k &=& \ln \Big{[}\left(\frac{k_y}{k_z}\right)^v \left(\frac{k}{k_2}\right)^{\beta - 1}\left(\frac{k_2}{k_1}\right)^{\beta} \left(\frac{k_s}{k_1}\right)^{-\beta_s} \Big{]},  ~~~ k_H \leqslant k \leqslant k_2, \label{rk5}
\end{eqnarray}
For the modes in the long wavelength regime $k\leqslant k_H$, one has to take $a(\varsigma_R)$, i.e., at reception, for $a_{\ast \ast} (k)$, the frequency corresponding to the conformal time at the starting time of the range. The rest of the calculations are same for the short wavelength modes. Then, the squeezing parameter for the rest of the frequency ranges are found to be
\begin{eqnarray}
r_k &=& \ln \Big{[}\left(\frac{k_y}{k_z}\right)^v \left(\frac{k}{k_H}\right)^{\beta + 1} \left(\frac{k_H}{k_2}\right)^{\beta - 1} \left(\frac{k_2}{k_1}\right)^{\beta}   \left(\frac{k_s}{k_1}\right)^{-\beta_s} \Big{]},~~~ k_E \leqslant k \leqslant k_H, \label{rk6}\\
r_k &=& \ln \Big{[}\left(\frac{k_y}{k_z}\right)^v \left(\frac{k}{k_H}\right)^{\beta} \left(\frac{k_E}{k_H}\right) \left(\frac{k_H}{k_2}\right)^{\beta - 1} \left(\frac{k_2}{k_1}\right)^{\beta}  \left(\frac{k_s}{k_1}\right)^{-\beta_s}\Big{]},~~~ k \leqslant k_E. \label{rk7}
\end{eqnarray}

\section{The spectrum of primordial gravitational waves} 
\label{sec4}
The amplitude of stochastic background of inflationary gravitational waves can be determined with eq.(\ref{hkbe}). In addition to the amplitude, the stochastic primordial gravitational waves can also be characterized by the spectral energy density $\Omega_{gw}$. The spectral energy density can be given in terms of the field as,
\begin{equation}\label{om}
\Omega_{gw}(\nu)=\frac{\pi^2}{3} h^2(\nu, \varsigma_H) \left(\frac{\nu}{\nu_H}\right)^2.
\end{equation}
The wave number $k$ is proportional to the frequency $\nu$, so the ratios of the wave numbers can be replaced by the ratios of the frequencies. The Hubble frequency is $\nu_H \equiv 1/l_H \simeq 2 \times 10^{-18}$ Hz. For other values of frequency, we choose $\nu_E = 1.5 \times 10^{-18} $ Hz which is in the long wavelength regime, $\nu_2 = 117 \times 10^{-18}$ Hz and $\nu_s = 10^8$ Hz. The $e^+e^-$ annihilation takes place around the frequency $10^{-12}-10^{-10}$ Hz, therefore, we used $\nu_y = 10^{-10}$ Hz and $\nu_z = 10^{-12}$ Hz \cite{swy}.

The frequency $\nu_1$ corresponds to the highest frequency at which the spectral energy density $\Omega_{gw} (\nu)$ in high frequency modes does not exceed the nucleosynthesis bound. Since $r_k = 0$ for $k_1$, then from eq.\eqref{hkbe}, we have,
\begin{equation}\label{nu1}
h(\nu_1) = 8\sqrt{\pi}\left(\frac{l_{pl}}{l_H}\right)\left(\frac{\nu_1}{\nu_H}\right).
\end{equation}
The big bang nucleosynthesis bound gives the upper limit on energy density $\Omega_{gw} < 1.5 \times 10^{-5}$ \cite{bbn4}, which leads to,
\begin{equation}
\Omega_{gw} h_0^2 < 6.855 \times 10^{-6},
\end{equation}
where $h_0 \sim 0.676$ is the reduced Hubble parameter \cite{plk}. Then, assuming $\Omega_{gw} (\nu_1) \approx 10^{-6}$, and using $l_H/l_{pl} = 9.276 \times 10^{59}$ and eq.\eqref{nu1} in eq.\eqref{om}, we get $\nu_1 = 1.2 \times 10^{10}$ Hz.

The initial amplitude of the primordial gravitational waves can be obtained from eq.\eqref{a1} as \cite{lp1, yz1},
\begin{equation}
h(k,\varsigma_i)=A\left(\frac{k}{k_H}\right)^{2+\beta},
\end{equation}
where $A=8\sqrt{\pi}bl_{pl}/l_0$. For the normalization factor $A$, it has been assumed that the CMB anisotropies at low multipoles are induced by the gravitational waves such that $\Delta T/T \simeq h(k, \varsigma_H) = A\frac{1}{(1+z_E)^3}$ \cite{yz1}. Further, $1/(1+z_E)^3=\Omega_m/ \Omega_\Lambda$ and we have considered the dark matter $\Omega_m \sim 0.3$ and dark energy $\Omega_\Lambda \sim 0.7$. 

From the BB-mode power spectrum for the 2018 BICEP/Keck Array (at 150 GHz and 220 GHz) and Planck (at 353 GHz) collaboration, the angular power spectrum $\ell (\ell + 1)C_{\ell}/2\pi$ at $\ell \sim 80$ are $\sim 0.28 ~\mu {\rm K^2}$ for BK18$_{150}\times $P$_{353}$ and $\sim 0.8 ~\mu {\rm K^2}$ for BK18$_{220}\times $P$_{353}$. This leads to $\Delta T/T \approx 9.7 \times 10^{-7}$ for BK18$_{150}\times $P$_{353}$ and $\Delta T/T \approx 1.64 \times 10^{-6}$ for BK18$_{220}\times $P$_{353}$. Then, from these with the above assumption, we get the normalization factor $A \approx 3 \times 10^{-6}$. This  eventually leads to $bl_{pl}/l_0=2.11 \times 10^{-7}$. 

\subsection{Inflationary models}
From the power spectrum of primordial gravitational waves, the inflationary index $\beta$ can be related to the tensor spectral index as $n_t = 2\beta + 4$. The Planck and BICEP/Keck Array collaboration constraint on the tensor-to-scalar ratio $r_{0.05}=0.014^{+0.010}_{-0.011}$ provides the upper limit as $r \leqslant 0.036$ \cite{bkp}. This limit rules out many inflation models. Among the models which are in good agreement with this result are the Starobinsky model ($R^2$ model) and the $\alpha$-attractor models \cite{as4, alp1, alp2, alp3}. 

The $\alpha$-attractor mechanism predicts flat potential which is the ideal condition for slow-roll inflation. The T-models of $\alpha$-attractor models have the potential,
\begin{equation}
V = V_0 \tanh^2\frac{\varphi}{\sqrt{6\alpha}},
\end{equation}
where $V_0$ is the height of the potential and models with sufficiently small values of $\alpha$ are consistent with observational data.

At large e-folding number $N$, the predictions of these models for the scalar spectral index $n_s$ and the tensor-to-scalar ratio $r$ depend only on $V_0$ and $\alpha$ and are given by,
\begin{eqnarray}
n_s &=& 1-\frac{2}{N}, \nonumber \\
r &=& \frac{12\alpha}{N^2}.
\end{eqnarray}
The E-models of the $\alpha$-attractor models have the potential,
\begin{equation}
V=V_0 \left(1-e^{-\sqrt{2/3\alpha}\varphi}\right)^2.
\end{equation}
For $\alpha = 1$, the potential for these models coincides with that of the $R^2$ model. In small $\alpha$ limit, the predictions of both the T- and E-models coincide with each other and are consistent with the BKP18 bound for $\alpha \lesssim 7$.

Here, we shall use $\alpha = 1$ and $\alpha = 7$ both for $N=50$ and $N=60$ for demonstration where,
\begin{eqnarray}
r &=& 4.8 \times 10^{-3},~~~~{\rm for}~\alpha = 1,~ N=50; \nonumber \\
r &=& 3.36 \times 10^{-2},~~~{\rm for}~\alpha = 7,~ N=50; \nonumber \\
r &=& 3.33 \times 10^{-3},~~~{\rm for}~\alpha = 1,~ N=60;  \\
r &=& 2.33 \times 10^{-2},~~~{\rm for}~\alpha = 7,~ N=60. \nonumber
\end{eqnarray}
Then, we have,
\begin{eqnarray}
\beta &=& -2.0003,~~~~{\rm for}~\alpha = 1,~ N=50; \nonumber \\
\beta &=& -2.0021,~~~~{\rm for}~\alpha = 7,~ N=50; \nonumber \\
\beta &=& -2.00021,~~~{\rm for}~\alpha = 1,~ N=60;  \\
\beta &=& -2.00146,~~~{\rm for}~\alpha = 7,~ N=60. \nonumber
\end{eqnarray}
Using the values of $bl_{pl}/l_0$ and eq.\eqref{vs2} with the ratios of $k$'s corresponding to those of $\nu$'s in eq.\eqref{ars}, we get the values of $\beta_s$ for each $\beta$ as:
\begin{eqnarray}\label{bs_ne}
\beta_s &=& -2.18366,~~~{\rm for}~\alpha = 1,~ N=50; \nonumber \\
\beta_s &=& -2.20771,~~~{\rm for}~\alpha = 7,~ N=50; \nonumber \\
\beta_s &=& -2.18246,~~~{\rm for}~\alpha = 1,~ N=60;  \\
\beta_s &=& -2.19916,~~~{\rm for}~\alpha = 7,~ N=60. \nonumber
\end{eqnarray}
The evaluation for the set of eqs.\eqref{bs_ne} is done when $e^+e^-$ annihilation stage is included in the evolutionary stage. Since the expressions for eqs.\eqref{vrs}-\eqref{ars} slightly differ with and without the inclusion of the $e^+e^-$ annihilation stage \cite{yz1, np}, it is clear that the subsequent evaluations for $\beta_s$ for each $\beta$ will also differ (see Appendix \ref{appa}). Thus, when the $e^+e^-$ annihilation stage is not included in the overall spectrum, which we used for comparison, the same evaluation leads to,
\begin{eqnarray}
\beta_s &=& -2.12441,~~~{\rm for}~\alpha = 1,~ N=50; \nonumber \\
\beta_s &=& -2.14846,~~~{\rm for}~\alpha = 7,~ N=50; \nonumber \\
\beta_s &=& -2.1232,~~~~{\rm for}~\alpha = 1,~ N=60;  \\
\beta_s &=& -2.13991,~~~{\rm for}~\alpha = 7,~ N=60. \nonumber
\end{eqnarray}

\begin{figure*}
\centering
\subfloat[]
{\includegraphics[scale=0.37]{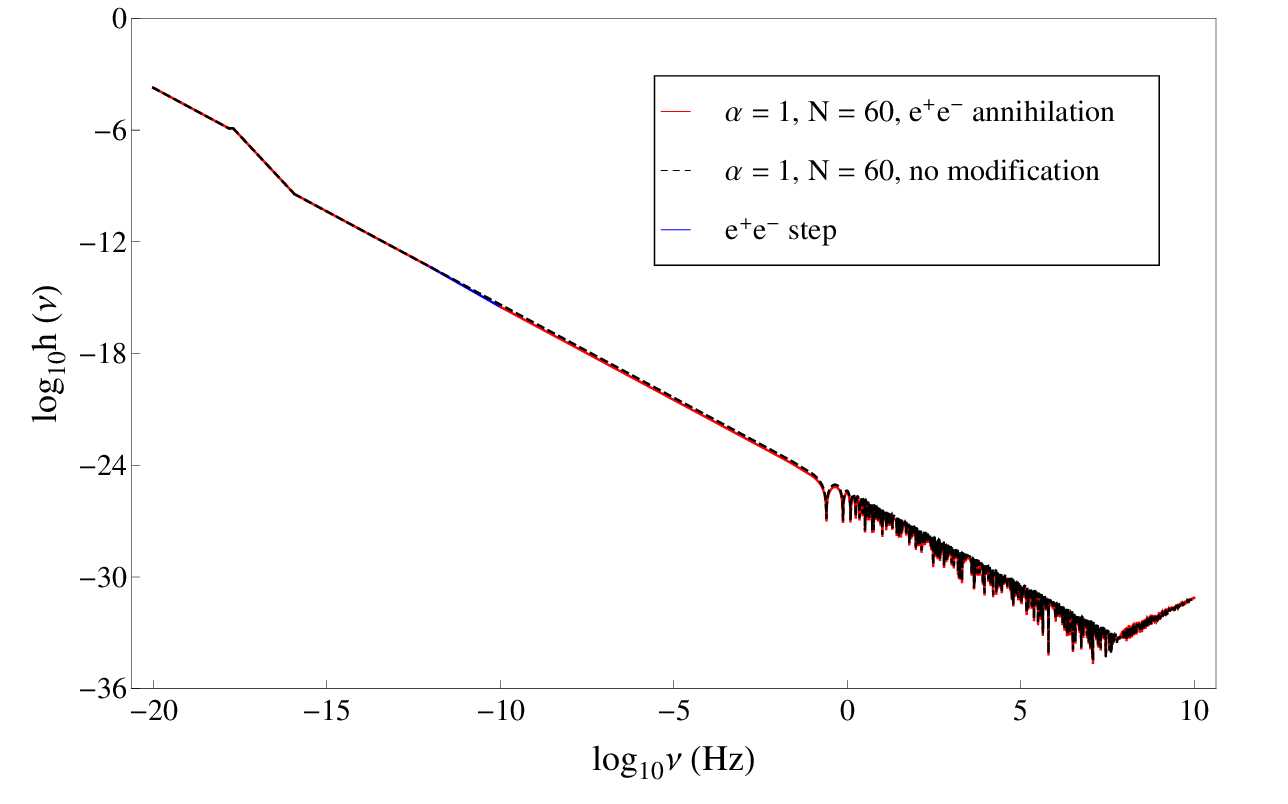}
\label{spec}}
\subfloat[]
{\includegraphics[scale=0.37]{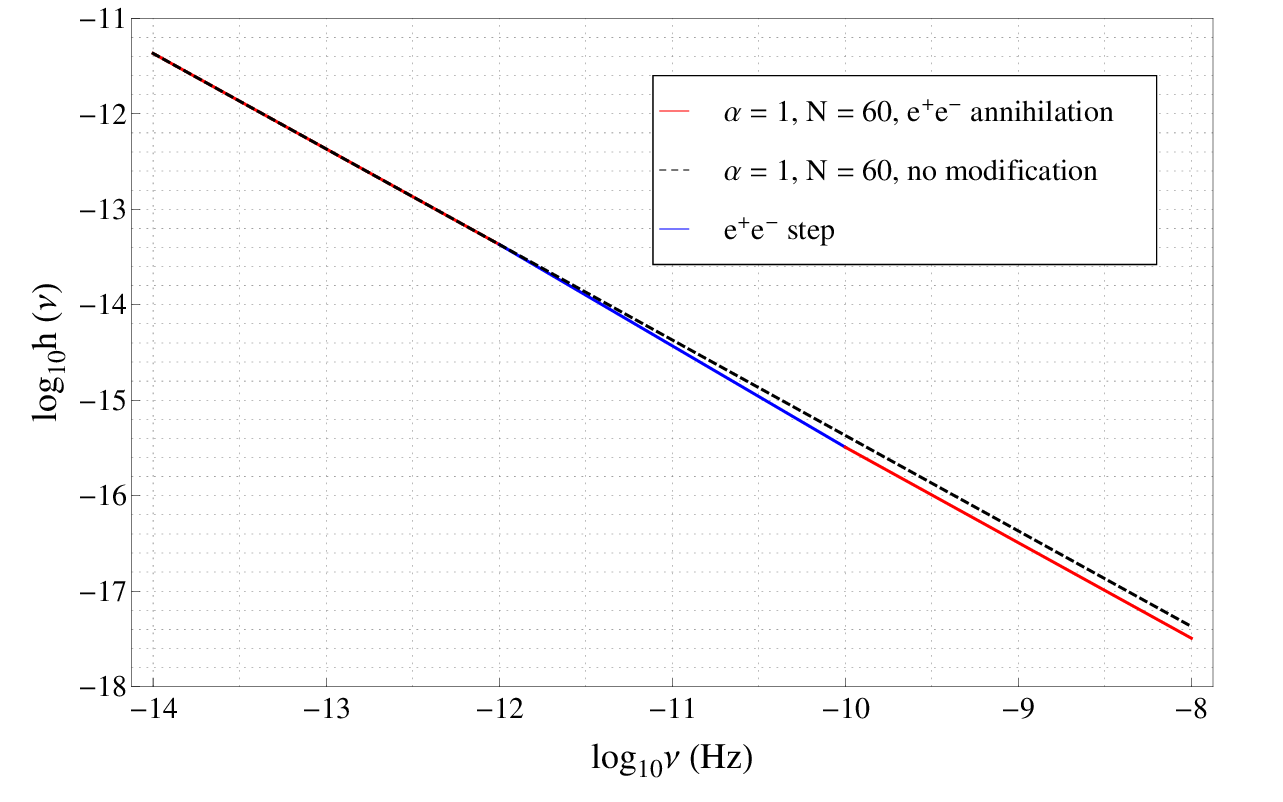}
\label{spec_s}}
\caption{Amplitude for $R^2$ inflationary model ($\alpha=1$) with $N=60$. \ref{spec_s} shows a close-up look into \ref{spec} for $10^{-14} \leqslant \nu \leqslant 10^{-8}$ Hz. There is a step of $\sim 10\%$ which starts at $\nu \geqslant 10^{-12}$ Hz due to $e^+e^-$ annihilation.}
\label{amps}
\end{figure*}

\begin{figure*}
\centering
\subfloat[]
{\includegraphics[scale=0.37]{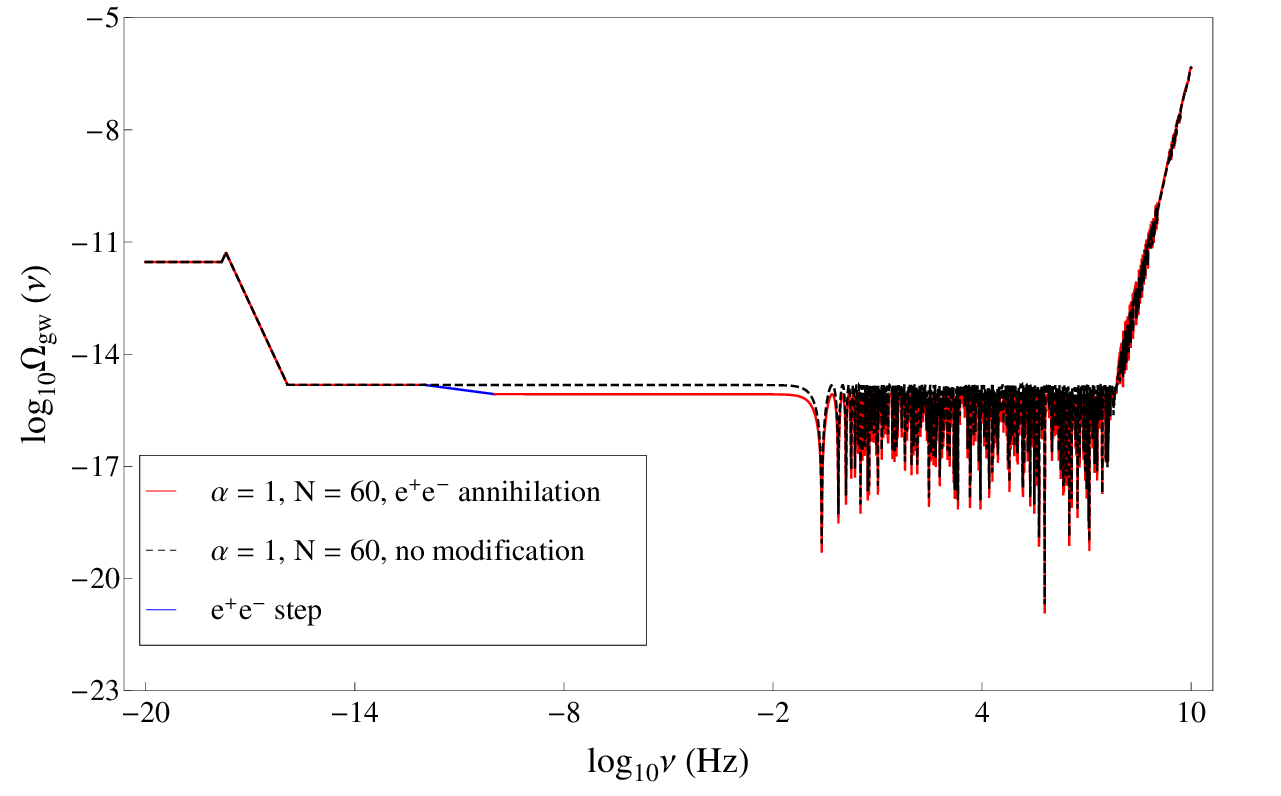}
\label{energy}}
\subfloat[]
{\includegraphics[scale=0.37]{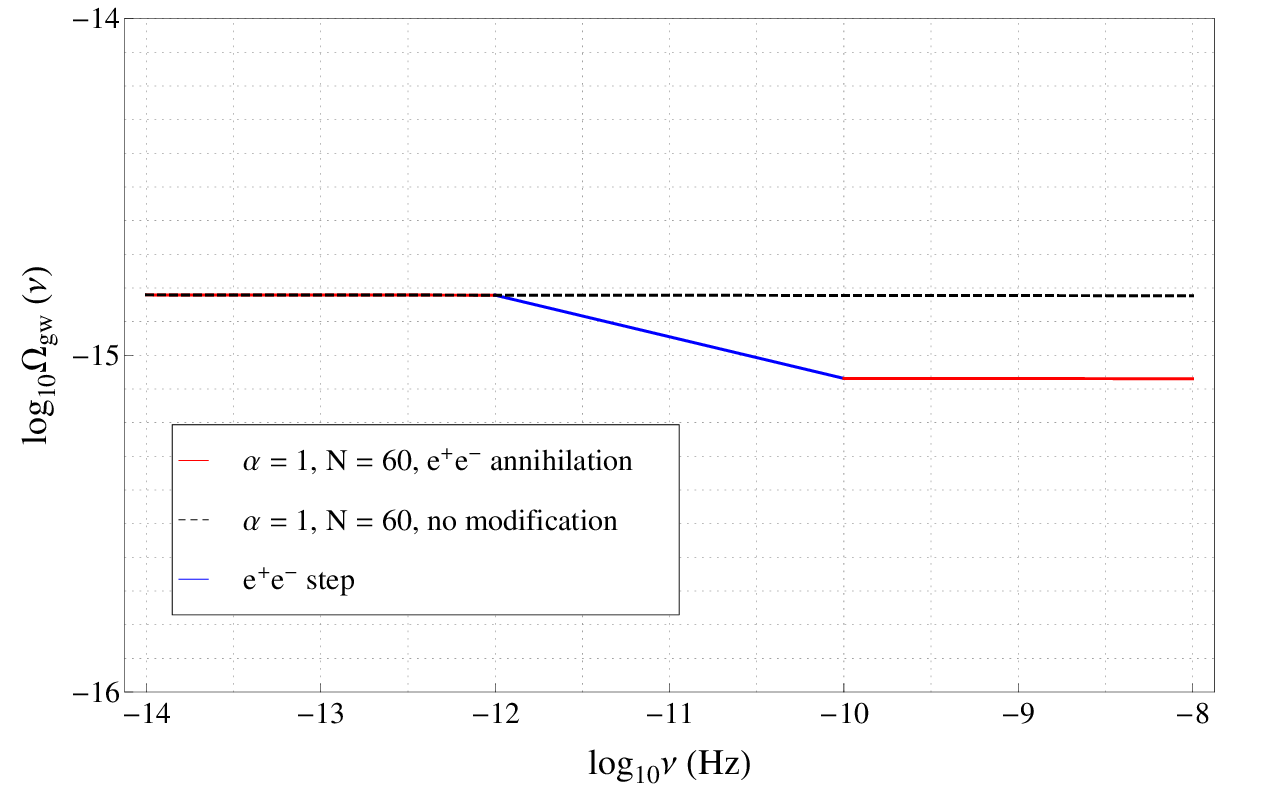}
\label{energy_s}}
\caption{Spectral energy density for $R^2$ inflationary model ($\alpha=1$) with $N=60$. \ref{energy_s} shows a close-up look into \ref{energy} for $10^{-14} \leqslant \nu \leqslant 10^{-8}$ Hz. There is a step of $\sim 25\%$ which starts at $\nu \geqslant 10^{-12}$ Hz due to $e^+e^-$ annihilation.}
\label{ens}
\end{figure*}
In figures \ref{amps}, we show the amplitude for the model $\alpha=1$, $N=60$. Figure \ref{spec} shows the amplitude over all frequencies while \ref{spec_s} is a zoom in to show the step in the spectrum due to $e^+e^-$ annihilation which is about $\sim 10\%$ as compared to the spectrum without $e^+e^-$ annihilation (labeled no modification). The grid lines are included in figure \ref{spec_s} to estimate the amount of the step correction and each horizontal grid line accounts for $20\%$ difference with its immediate neighbor. For the former, we have used $\beta_s = -2.18246$ and the solid red line with blue insert which indicates the $e^+e^-$ step and for the latter, we have used $\beta_s = -2.1232$ and dashed black lines.

Figures \ref{ens} show the spectral energy density for $\alpha=1$, $N=60$. Color assignment is the same as those for the amplitude plots and each horizontal grid line accounts for $10\%$ difference. Figure \ref{energy} shows the spectral energy density over all frequencies and \ref{energy_s} is a zoom in into \ref{energy} which shows the step in the spectrum due to $e^+e^-$ annihilation which is about $\sim 25\%$ as compared to the spectrum without $e^+e^-$ annihilation (labeled no modification). Repeating the same evaluations for amplitude and energy density for other $\alpha$ models leads to the same results.

\begin{figure*}
\centering
\subfloat[]
{\includegraphics[scale=0.37]{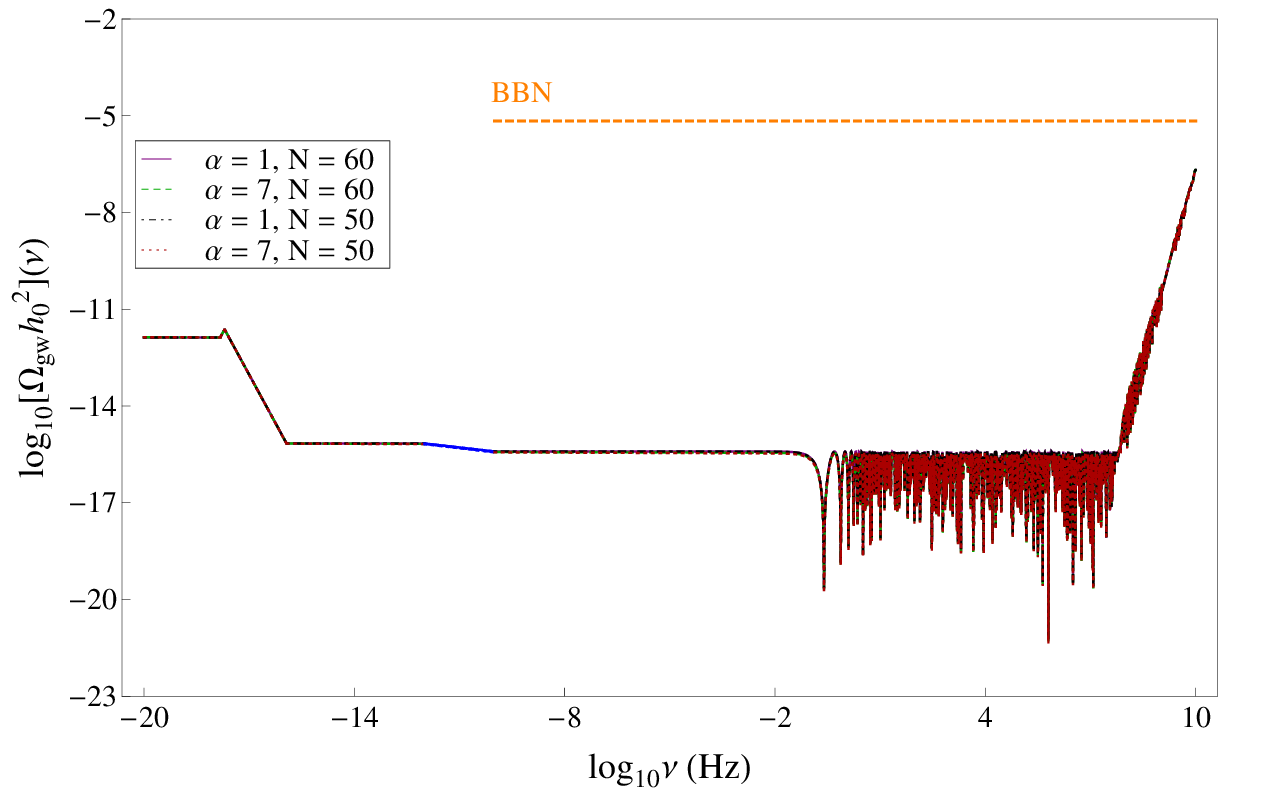}
\label{bbn}}
\subfloat[]
{\includegraphics[scale=0.37]{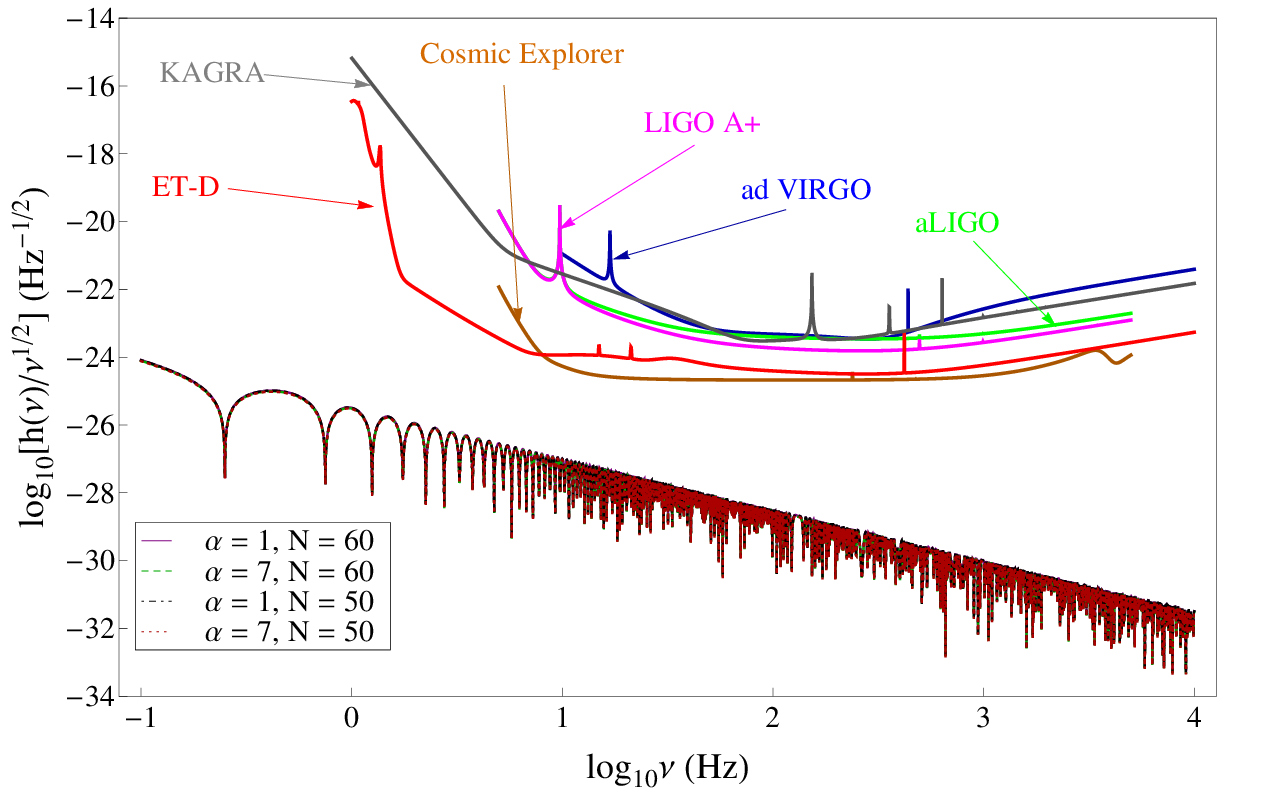}
\label{ampx}}
\caption{Energy densities for $\alpha$-attractor models tested with BBN bound and amplitudes with sensitivity curves of various detectors.}
\label{bbps}
\end{figure*}
In figures \ref{bbps}, we show the energy density $\Omega_{gw}h_0^2$ tested with the BBN bound (dashed orange) and the root mean square amplitude $h(\nu)\nu^{-1/2}$ for the models $\alpha=1$, $\alpha=7$, both for $N=50$ and $N=60$ with the $e^+e^-$ annihilation stage tested with the sensitivity curves of the currently ongoing and upcoming detectors - advanced LIGO (light green), advanced Virgo (blue), LIGO A+ (magenta), KAGRA (dark grey), Einstein Telescope (red) and Cosmic Explorer (dark orange) which operate in the frequency range $\mathcal{O}(10-10^4)$ Hz and can probe upto amplitudes of the order of $10^{-24}-10^{-25}$. In both figures \ref{bbn} and \ref{ampx}, the models $\alpha=1$, $N=60$; $\alpha=7$, $N=60$; $\alpha=1$, $N=50$; and $\alpha=7$, $N=50$ are shown with purple, dashed green, dot-dashed black and dotted red plots respectively.

It can be seen from figure \ref{bbn} that the spectra for all the models demonstrated are extremely close to one another. All of them are well within the BBN bound. However, figure \ref{ampx} shows that the field for these models are not within reach of the current and upcoming sensitivity curves of the detectors.

\section{Conclusions}\label{sec5}
In this paper, we studied the stochastic background of inflationary gravitational waves modified by the $e^+e^-$ annihilation and tested $\alpha$-attractor inflationary models highly favored by BKP18 data. We used the models $\alpha=1$, $N=60$; $\alpha=7$, $N=60$; $\alpha=1$, $N=50$; and $\alpha=7$, $N=50$ for demonstration of the spectra. All these models normalized with the BKP18 data are well within the BBN bound, but are very much below the sensitivity curves of the currently operating and proposed detectors. Future CMB missions are also expected to observe and implement tighter and lower constraint on the tensor-to-scalar ratio indicating the requirement of extremely sensitive detectors for a speck of hope for direct detection of primordial gravitational waves. Therefore, unless the B-modes are detected, the direct detection of primordial gravitational waves are very much out of question in the present and near future.

\appendix
\numberwithin{equation}{section}
\section{Particle physics before and after $e^+e^-$ annihilation}
\label{app}
During the radiation dominated era, the particles which interact with the photon are in thermal equilibrium. Entropy per unit comoving volume is conserved in the adiabatic system,
\begin{eqnarray}
S(T) &=& s(T) a^3 (T) = {\rm constant}, \label{ap1}\\
{\rm for}~s(T) &=& (\rho_r + p_r)/T = g_{\ast s} (T) \frac{2\pi^2}{45} T^3 \label{ap2}
\end{eqnarray}
where $s(T)$ is the entropy density and the energy density $\rho_r(T)$ and pressure $p_r (T)$ are, 
\begin{eqnarray}
\rho_r(T) &=& g_\ast (T) \frac{\pi^2}{30}T^4, \label{ap3} \\
p_r (T) &=& \frac{1}{3}\rho_r (T),
\end{eqnarray}
respectively, $g_{\ast s}$ and $g_{\ast}$ being the effective number of relativistic species contributing to the entropy and energy density respectively given by
\begin{eqnarray}
g_{\ast s} &=& \sum_{i={\rm boson}} g_i \left(\frac{T_i}{T}\right)^3 + \frac{7}{8} \sum_{i={\rm fermion}}g_i \left(\frac{T_i}{T}\right)^3, \label{ag} \\
g_{\ast} &=& \sum_{i={\rm boson}} g_i \left(\frac{T_i}{T}\right)^4 + \frac{7}{8} \sum_{i={\rm fermion}}g_i \left(\frac{T_i}{T}\right)^4, \label{ags}
\end{eqnarray}
where $g_i$ is the number of spin or helicity states of $i$-th species, $T_i$ its temperature and $T$ for photon temperature $T_\gamma$. $g_\ast(T)$ and $g_{\ast s} (T)$ are equal for temperature $T \gtrsim 0.1$ MeV. Then, using eqs.\eqref{ap1} and \eqref{ap2} in eq.\eqref{ap3}, we get
\begin{equation}\label{ap4}
\rho_r \propto g_\ast g_{\ast s}^{-4/3} a^{-4}.
\end{equation}
Just before $e^+e^-$ annihilation, all particles were in equilibrium at the same temperature such that $g_\ast (T) = g_{\ast s} T$ and the plasma consists of photons (2-helicity states), electrons and positrons (2 spin states each) and three species each of neutrino and antineutrino (1-helicity state each) all at the same temperature $T_1$, say, then
\begin{equation}\label{ap5}
s = \frac{2\pi^2}{45} T_1^3 \left[2 + \frac{7}{8}(2+2+6)\right].
\end{equation}
Then, from eqs.\eqref{ap2} and \eqref{ap5}, we get
\begin{equation}
g_\ast = g_{\ast s} = 10.75.
\end{equation}
$e^+e^-$ annihilation starts around $T \sim 0.5$ MeV and ends around $T \sim 0.1$ MeV \cite{ywk,swy}. Just after $e^+e^-$ annihilation and while the three neutrino species remain relativistic, we have only photons and three families each of neutrinos and antineutrinos with two different temperatures $T_\gamma$ and $T_\nu$ respectively since at temperatures below the $e^+e^-$ annihilation temperature, neutrinos are cooler than photons which gain heat from the $e^+e^-$ annihilation, and
\begin{equation}\label{ap6}
s = \frac{2\pi^2}{45} \left[2 T_\gamma^3 + \frac{7}{8}\times 6 T_\nu^3 \right],
\end{equation}
where $T_\nu = (4/11)^{1/3}T_\gamma$. Then, using eqs.\eqref{ag}, \eqref{ags} and \eqref{ap6}, we get the values of $g_{\ast s}$ and $g_{\ast}$ after $e^+e^-$ annihilation but before the heaviest neutrino becomes non-relativistic as,
\begin{eqnarray}
g_{\ast s} &=& 3.9091, \\
g_\ast &=& 3.363.
\end{eqnarray}
Now, from Friedmann equation, during radiation domination, we have,
\begin{equation}
\left(\frac{a'}{a^2}\right)^2 = \frac{8\pi G}{3} \rho_r.
\end{equation}
Then, using eq.\eqref{ap4}, we get
\begin{equation}\label{ap7}
a' = g_\ast^{1/2} g_{\ast s}^{-2/3}.
\end{equation}
Taking the conformal time rate of expansion $a'$ before and after the $e^+e^-$ annihilation from eqs.\eqref{a3} and \eqref{a5} and using them in eq.\eqref{a7}, we get the ratio
\begin{equation}\label{ap8}
\frac{a_g}{a_e} = \frac{g_\ast^{-1/2}(\varsigma_y)g_{\ast s}^{2/3}(\varsigma_y)}{g_\ast^{-1/2}(\varsigma_z)g_{\ast s}^{2/3}(\varsigma_z)} \simeq 1.098.
\end{equation}
Also, from eqs.\eqref{ars}, we get the ratio
\begin{equation}\label{ap9}
\frac{a_g}{a_e} = \xi_y^{v/1+v}.
\end{equation}
The $e^+e^-$ annihilation takes place during the temperatures $T \sim (0.5\sim 0.1)$ MeV, and since $T \propto 1/a(\varsigma)$, we have
\begin{equation}
\xi_y = \frac{a(\varsigma_z)}{a(\varsigma_y)} = 5.
\end{equation}
Using this with eqs.\eqref{ap8} and \eqref{ap9} yields $v \approx 0.0616$, the index that characterizes the $e^+e^-$ annihilation stage.

\section{Growth of mode $h_k(\varsigma)$}
\label{appb}
Using eqs.\eqref{mff} and \eqref{y} in \eqref{mf}, we can write
\begin{eqnarray}
h_k(\varsigma)a(\varsigma) &=& e^{i \theta_k} \cosh r_k + e^{i(\theta_k -2\phi_k)} \sinh r_k \nonumber \\
|h_k(\varsigma)|^2 a^2 (\varsigma) &=& \left(e^{i \theta_k} \cosh r_k + e^{i(\theta_k -2\phi_k)} \sinh r_k\right)  \times \left( e^{-i \theta_k} \cosh r_k + e^{-i(\theta_k -2\phi_k)} \sinh r_k \right) \nonumber \\
 &=& \cosh 2r_k + 2\sinh 2r_k \cos 2\phi_k,
\end{eqnarray}
which is precisely eq.\eqref{axx}. Now, we can expand this equation,
\begin{eqnarray}
|h_k(\varsigma)|^2 a^2 (\varsigma) &=& \cosh 2r_k (\cos^2 \phi_k + \sin^2 \phi_k )  + 2\sinh 2r_k (\cos^2 \phi_k - \sin^2 \phi_k ) \nonumber \\
 &=& \cos^2 \phi_k (\cosh 2r_k + \sinh 2r_k)  + \sin^2 \phi_k (\cosh 2r_k - \sinh 2r_k) \nonumber \\
 &=& \cos^2 \phi_k e^{2r_k} + \sin^2 \phi_k e^{-2r_k}. \label{gg}
\end{eqnarray}
As squeezing parameter $r_k$ grows with time, i.e., decrease in frequency, we have $r_k \gg 1$ as it grows, hence we can neglect the second term in the right hand side of eq.\eqref{gg}. Therefore, we can write
\begin{equation}
h_k (\varsigma) a(\varsigma) \approx e^{r_k} \cos \phi_k.
\end{equation}
Hence, with the evolution of the universe, the mode grows such that $h_k (\varsigma) = h_k^\ast (\varsigma)$, the complex function becomes practically real.

\section{Evaluation for $\beta_s$ without $e^+e^-$ annihilation}
\label{appa}
When the $e^+e^-$ annihilation stage is not considered explicitly in the evolutionary course of the universe, eq.\eqref{ars} can be written as \cite{lp1, yz1},
\begin{equation}
l_0 = l_H b\xi_E^{-(2+\beta)}\xi_2^{(\beta-1)/2}\xi_s^{\beta}\xi_1^{(\beta-\beta_s)/(1+\beta_s)}, \label{lo}
\end{equation}
where,
\begin{eqnarray}\label{vs3}
&& \xi_1 = \left(\frac{k_1}{k_s}\right)^{1+\beta_s},~~~~~~~\xi_s = \frac{k_s}{k_2}, \nonumber \\
&& \xi_2 = \left(\frac{k_2}{k_E}\right)^2, ~~~~~~~~~~\xi_E = \left(\frac{k_E}{k_H}\right)^{-1}.
\end{eqnarray}
Then, using the same normalization and following in the same steps of using the values of $bl_{pl}/l_0=2.11 \times 10^{-7}$ and eq.\eqref{vs3} with the ratios of $k$'s corresponding to those of $\nu$'s in eq.\eqref{lo}, we can get $\beta_s$ values for each model without the $e^+e^-$ annihilation.

\end{document}